\documentclass[
 reprint,
 amsmath,amssymb,
 aps,
]{revtex4-2}
\usepackage{hyperref}
\usepackage{graphicx}% Include figure files
\usepackage{dcolumn}% Align table columns on decimal point
\usepackage{bm}% bold math

\begin{document}

\preprint{APS/123-QED}

\title{Effects of Multi-scale Coupling on Particle Acceleration and Energy Partition in Magnetic Reconnection}

\author{A. Velberg}
\email{velberg@mit.edu}
\author{N.F. Loureiro}
 
\affiliation{
 Plasma Science and Fusion Center, Massachusetts Institute of Technology, Cambridge, MA 02139, USA
}

\author{A. Stanier, X. Li, F. Guo, and W. Daughton}
\affiliation{
 Los Alamos National Laboratory, Los Alamos, NM 87545, USA
}%

\date{\today}

\begin{abstract}

The interplay between kinetic and macroscopic scales during magnetic reconnection is investigated using particle-in-cell simulations of magnetic island coalescence in the strongly-magnetized, relativistic pair plasma regime. For large system sizes, secondary current sheet formation and downstream turbulence driven by the reconnection outflows dominate the global energy dissipation so that it is causally connected, but spatially and temporally de-coupled from the primary reconnecting current sheet. When compared to simulations of an isolated, force-free current sheet, these dynamics activate additional particle acceleration channels which are responsible for a significant population of the non-thermal particles, modifying the particle energy spectra. 
\end{abstract}

\maketitle

\textit{Introduction.}---Magnetic reconnection is frequently invoked to explain the explosive release of energy and high-energy emissions observed in a variety of astrophysical plasma environments \cite{sironi_relativistic_2014,dahlin_mechanisms_2014,li_acceleration_2021,guo_magnetic_2024,sironi_relativistic_2025}. As the associated re-configuration of magnetic field lines occurs in localized regions,  reconnection fundamentally involves the multi-scale coupling between kinetic-scale microphysics and larger-scale structures. Despite this, the consequences of this cross-scale feedback for the partition of released energy and mechanisms of particle acceleration remain poorly understood. 

Most of the previous studies of energy partition and particle acceleration focus on the microphysics local to the reconnecting current sheet \cite{li_acceleration_2021,ji_magnetic_2022,guo_magnetic_2024}. The paradigmatic setup involves an isolated, pre-formed current layer which is super-critically plasmoid-unstable. Many simulation studies using such a setup have characterized the partition of energy and mechanisms of particle acceleration, demonstrating the ability of reconnection to produce hard power-law tails in particle energy spectra \cite{li_acceleration_2021,sironi_relativistic_2014,guo_recent_2020,hoshino_efficiency_2022,hoshino_energy_2023,guo_magnetic_2024,sironi_relativistic_2025}. 

In nature, however, the current sheet forms dynamically and may involve the two-way feedback between kinetic scales and the larger-scale structures supplying the magnetic flux \cite{ji_magnetic_2022}. For example, reconnection onset in solar flares may be triggered by a macroscale instability \cite{hood_kink_1979,kliem_torus_2006,shibata_solar_2011}, playing a crucial role in the energy release and global system evolution. Large-scale feedback provided by the reconnection outflows may also play an important role in these processes. Recent observations from turbulent outflow regions driven by solar flare current sheets \cite{kontar_turbulent_2017,fleishman_solar_2022} and magnetotail reconnection \cite{angelopoulos_electromagnetic_2013,richard_turbulence_2024} suggest that sites of strong energy conversion and non-thermal particle acceleration due to reconnection are not limited to the local extent of the current sheet. 

In this Letter, we demonstrate that the multi-scale coupling between the current sheet and global system is essential to understanding the details of energy partition and particle acceleration in reconnecting systems. To investigate these effects, we perform fully-kinetic particle-in-cell (PIC) simulations of large-scale coalescing magnetic islands, which include self-consistent current sheet formation due to the coalescence instability \cite{finn_coalescence_nodate,pritchett_coalescence_1992}, and fully-resolved outflow regions. Remarkably, we find that for large system sizes,  secondary current sheets and turbulence driven by the outflows become the dominant sites of energy transfer and particle acceleration, allowing these processes to be both spatially and temporally decoupled from the dynamics at the local reconnection site. Although the primary current sheet remains essential to accelerating particles to the highest energies, additional acceleration mechanisms downstream provide an important channel for energy release which accelerates a large population of particles to non-thermal energies. 

\textit{Numerical setup.}---Simulations are performed using the fully-kinetic PIC code VPIC \cite{bird_vpic_2022}. The island coalescence initial condition is a force-free modification of the Fadeev \cite{fadeev_self-focusing_1965} equilibrium, with magnetic field components 

\begin{align}
    B_x &= B_0 \sinh(x/L)/ F(x,z), \nonumber\\
    B_y &= B_0 \left ( \frac{1-\epsilon^2}{F(x,z)^2} \right )^{1/2},\label{eq:equil} \\
    B_z &= B_0\epsilon \sin(x/L)/F(x,z),\nonumber
\end{align}
where $B_0$ is the asymptotic field strength, $L$ is the island half-width in terms of $d_e$, $\epsilon=0.4$, and $F(x,z)= \cosh(z/L)+\epsilon\cos(x/L)$. There is an intrinsic order unity out-of-plane component ($B_y/B_0\sim\mathcal{O}(1)$), but no external guide field. A strongly-magnetized, relativistic electron-positron plasma is initialized with a Maxwell-Jüttner (M-J) distribution with spatially uniform density $n_0$ and thermal temperature $k_BT_i=k_BT_e = 0.36m_ec^2$. Particles are given a net drift velocity $\mathbf{U}_i=-\mathbf{U}_e$ consistent with $\nabla\times \mathbf{B} = 4\pi\mathbf{J}/c$, where the current density $\mathbf{J}=en_0(\mathbf{U}_i-\mathbf{U}_e)$. The magnetization parameter is $\sigma = B^2/(4\pi n_em_ec^2)=(\Omega_{ce}/\omega_{pe})^2=25$, where $\Omega_{ce}=eB/(m_ec)$ is the electron cyclotron frequency and $\omega_{pe}= \sqrt{4\pi n_e e^2/m_e}$ is the electron plasma frequency. The key parameter in this study is $L/d_e$, a proxy for macroscopic magnetic island size which is varied by choosing $L \in [20,40,80,160,320]d_e$. For all simulations, the electron-positron skin depth $d_e=c/\omega_{pe}$ is resolved by four grid cells, $d_e= 4\Delta x$, with 150 particles-per-cell. The simulation domain is $x \in [-2\pi L,2\pi L]$, $z \in [-\pi L,\pi L]$, which for the largest simulation corresponds to a grid size $16384\times8192$ and a total of 40B particles. The coalescence instability is triggered with a sinusoidal initial perturbation that pinches the island end-points with strength $\delta B = 0.1 B_0$. The boundary conditions are periodic in the $x$ direction and reflecting (conducting) for particles (fields) in the $z$ direction.

\begin{figure}[h]
\includegraphics{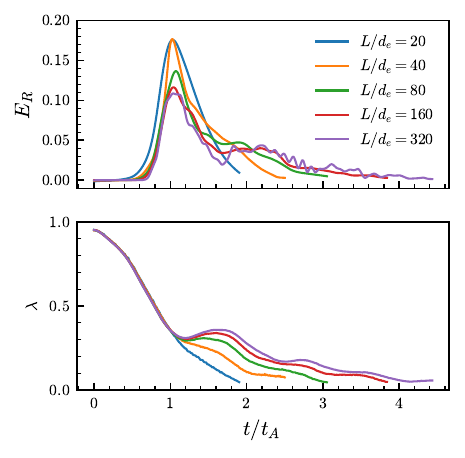}
\caption{\label{fig:dynamics} (Top) Reconnection rate $E_R$ versus time for island coalescence at several system sizes. (Bottom) Island O-point separation normalized to initial separation.}
\end{figure}

\textit{Island coalescence dynamics}.---Similar to previous studies using this setup \cite{knoll_coalescence_2006,karimabadi_flux_2011,stanier_role_2015,ng_island_2015}z, our simulations demonstrate characteristically ``bursty'' reconnection due to the coupling between the current sheet and large scale island motions. Figure \ref{fig:dynamics} shows the reconnection rate,  $E_R =-\partial_tA_{yX}/(B_mV_{Am})$ \cite{karimabadi_flux_2011,stanier_role_2015,ng_island_2015}, and separation between island O-points, $\lambda$, versus the global Alfvén time, $t/t_A = tV_{A}/(4\pi L) $, where $A_{yX}$ is the vector potential at the dynamically-identified X-point, $V_{A}=v_A/\sqrt{1+(v_A/c)^2}$ is the relativistic Alfvén speed with $v_A = B_0/\sqrt{4\pi n(m_i+m_e)}$, and $B_m$ and $V_{Am}$ are the maximum of initial reconnecting field component ($B_z$) between the islands and corresponding Alfvén speed. Data is shown up to $\tau_{99}$, the time at which $99\%$ of the initial flux in the islands has been reconnected for each run. 
The peak reconnection rates (see Table \ref{tab:table1}) exhibit a system size dependence which saturates at the typical value $E_R\sim 0.1$ \cite{cassak_review_2017,liu_why_2017} for large $L/d_e$ as reconnection transitions to the plasmoid-dominated regime \cite{ji_phase_2011}.

\begin{figure}[t]
    \centering
    \includegraphics[width=0.99\linewidth]{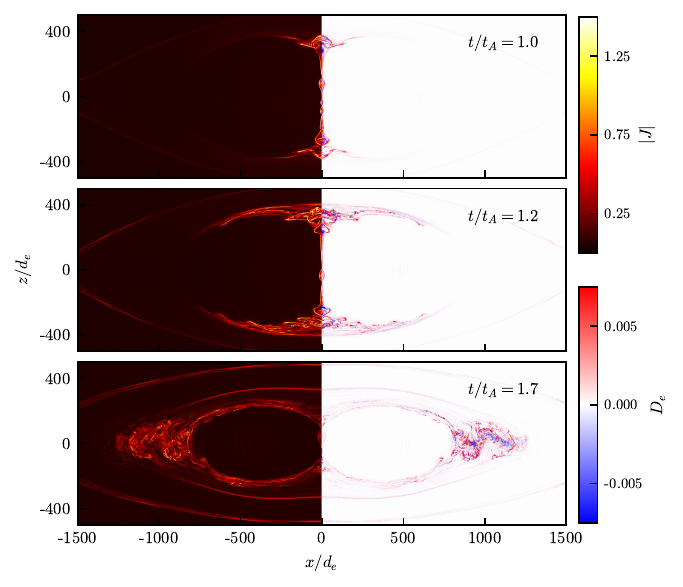}
    \caption{ Left side: Current density magnitude $|J|$ at several times for $L/d_e=320$ associated with: (top) $E_{R,\mathrm{max}}$, (center) first local minimum in $\lambda$, (bottom) $\lambda_{b1}$. Right side: $D_e$ at the same times.}
    \label{fig:temp2D}
\end{figure}

\begin{table}
\caption{\label{tab:table1}
Comparison of several parameters from island coalescence and FFCS simulations, including the maximum reconnection rate $E_{R,\mathrm{max}}$, maximum number of plasmoids in the system at a given time $N_p$, fraction of energy dissipated in the current sheet ($f_{\mathrm{CS}}$) and downstream region ($f_{\mathrm{DS}}$) at $\tau_{99}$, and energy band cutoffs $\gamma_\mathrm{nt}$ and $\gamma_\mathrm{hi}$ .
}
\begin{ruledtabular}
\begin{tabular}{lccccccc}
\textrm{$L/d_e$}&
\textrm{$E_{R,\mathrm{max}}$}&
\textrm{$N_p$}&
\textrm{$f_{\mathrm{CS}}$}&
\textrm{$f_{\mathrm{DS}}$}&
\textrm{$\gamma_\mathrm{nt}-1$}&
\textrm{$\gamma_\mathrm{hi}-1$}\\
\colrule
20 & 0.18 & 0 &0.54&0.31 & 5.23 & -- \\
40 & 0.18 & 3 &0.45&0.47 & 4.63 & 10.37 \\
80 & 0.14 & 8 &0.37&0.54 & 4.52 & 11.43 \\
160 & 0.12 &10 &0.32&0.62 & 4.56 & 10.87 \\
320 & 0.11 &23 &0.27&0.67 & 4.77 & 15.40 \\
\colrule

FFCS & 0.05 & 13 & -- & --  & -- & -- 
\end{tabular}
\end{ruledtabular}
\end{table}
For $L/d_e\le 40$, the current sheet is able to process all of the incoming flux from the islands. However, for  $L/d_e\geq80$, flux is supplied faster than it can be reconnected and begins to pile up upstream, creating a magnetic pressure force which causes the islands to bounce apart, as shown by the reversal in $\lambda$ near $t/t_A=1.2$. The current sheet is correspondingly attenuated and $E_R$ slows until the islands begin to coalesce again, leading to bursty reconnection rates. This evolution of the current sheet can be clearly observed in plots of the current density magnitude $|J|$ on the left side of Fig.~\ref{fig:temp2D}.  

These plots also highlight downstream dynamics driven by the reconnection outflows. Plasmoids formed in the current sheet (top panel) are ejected downstream, where they drive the formation of many secondary current sheets and small-scale fluctuations in broad ``wing'' regions (center panel). As outflows propagate further downstream, they converge with those from the opposite side of the current sheet, driving a large region of turbulence (bottom panel). Notably, because of the finite time for the outflows to arrive here, turbulence remains active during island bounces when the current sheet is weakened and $E_R$ is at a local minimum. As $L/d_e$ increases, increased scale separation leads to additional plasmoid activity (see Table \ref{tab:table1}) secondary current sheet formation, and outflow-driven turbulence, as can be observed in movies of $|J|$ at $L/d_e=40$ and 320, as well as a comparison of current density probability distribution functions for each run, found in the supplemental material. 

Bursty reconnection and outflow-driven turbulence are clear manifestations of the two-way feedback present in multi-scale reconnecting systems. As we will demonstrate, these dynamics have important consequences for the dominant energy partition channels and particle acceleration mechanisms in this system.

\textit{Energy dissipation.}---The downstream activity suggests that regions of strong energy conversion may not be exclusively localized to the primary reconnection site. To investigate this, we make use of the electron-frame dissipation measure \cite{zenitani_new_2011}, a proxy for collisionless energy dissipation in plasma turbulence and magnetic reconnection \cite{zenitani_new_2011,wan_intermittent_2015,pezzi_dissipation_2021}. It is given by \cite{zenitani_new_2011}
\begin{equation}
    D_e = \gamma [\mathbf{J} \cdot (\mathbf{E + V}_e \times \mathbf{B}) -\rho_c(\mathbf{V}_e\cdot \mathbf{E})], 
\end{equation}
where the Lorentz factor $\gamma\equiv 1/\sqrt{1-(V_e/c)^2}$, $V_e$ is the 3-velocity of the electron fluid, and $\rho_c$ is the charge density in the observer frame. As shown in the right half of the plots in Fig.~\ref{fig:temp2D}, regions of strong $D_e$ are co-located with the current sheets at the primary reconnection site as well as downstream. During the island bounce (bottom panel), secondary current sheets in the downstream turbulence are clearly the predominant sites of energy dissipation. 

To quantify the importance of different regions to the global energy dissipation, we have developed a robust method which uses particle tags to decompose the simulation domain into several regions at each time step, shown in different colors in the top panel of Fig.~\ref{fig:diss} (see supplemental material for details). The downstream (DS) and current sheet (CS) regions that are the focus of our analysis are shown in blue and orange, respectively. The cumulative $D_e$ integrated over each region, $\mathcal{D}_{i}=(1/M_0)\int_0^{t}dt'\int D_{e,i}(t') dV_i$, where $i$ is the region index and $M_0$ is the initial magnetic energy, is tracked for $L/d_e=40$ and 320 in the center and bottom panels of Fig.~\ref{fig:diss} up to $\tau_{99}$.  In the smaller system, the contribution from the primary current sheet ($\mathcal{D}_\mathrm{CS}$) rises rapidly following reconnection onset, before saturating as the island flux is reconnected. For $L/d_e=320$, however, the downstream region quickly takes over as the dominant location of energy dissipation, consistent with the development of turbulence in the wings and then further downstream. While the total fraction of magnetic energy dissipated is similar across systems,  $\mathcal{D}_\mathrm{CS}$ is notably smaller for $L/d_e=320$ due to lower reconnection rates and the mediation of the current sheet by bursty reconnection. The total fractional contributions to the global energy dissipation are summarized in Table~\ref{tab:table1}, showing that as $L/d_e$ increases, the dominant energy dissipation sites are increasingly de-localized from the primary current sheet.

\begin{figure}
    \centering
    \includegraphics[width=0.9\linewidth]{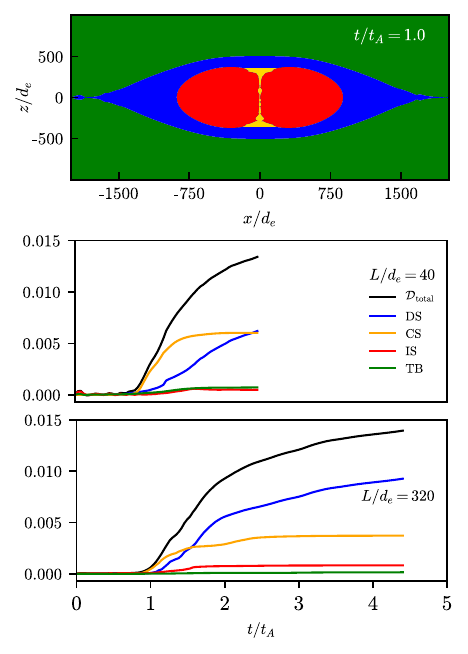}
    \caption{(Top) Region decomposition at time of $E_{R,\mathrm{max}}$ for $L/d_e=320$. Red: magnetic islands; orange: primary current sheet; blue: downstream region; green: top and bottom regions. (Center) Cumulative $\mathcal{D}_i$ normalized by initial magnetic energy, decomposed by region for $L/d_e=40$.  Colors match regions in top plot, with the system-wide total in black. (Bottom) Same as center for $L/d_e=320$.}
    \label{fig:diss}
\end{figure}

\begin{figure}
    \centering
\includegraphics[width=0.90\linewidth]{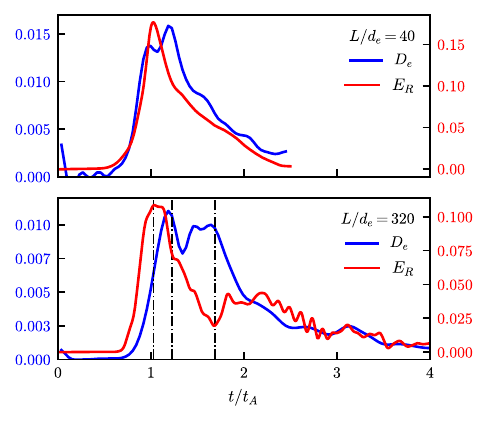}
    \caption{(Top) Volume integrated $D_e$ normalized to initial magnetic energy versus time for $L/d_e=40$ in blue, compared to $E_R$ in red, up to $\tau_{99}$. (Bottom) Same for $L/d_e=320$, with vertical dash-dotted lines indicating the times in Fig. \ref{fig:temp2D}.}
    \label{fig:diss_inst}
\end{figure}

A further consequence of this is to allow energy dissipation to remain strong despite the bursty reconnection rate. Figure \ref{fig:diss_inst} compares the global volume-integrated $D_e$ to the reconnection rate for the same simulations, confirming that while $D_e$ closely follows $E_R$ for $L/d_e=40$, activation of downstream dissipation sites allows it to remain high while $E_R$ is at a local minimum for $L/d_e=320$. The multi-scale coupling present in this system therefore allows strong energy dissipation to be both spatially de-localized and temporally de-coupled from the local reconnection dynamics, consistent with previous findings \cite{loureiro_fast_2013}. 

\begin{figure}
    \centering
    \includegraphics[width=0.98\linewidth]{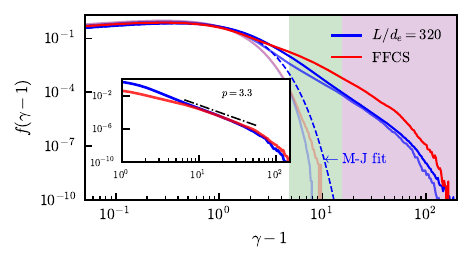}
    \caption{Electron energy spectra $f(\gamma-1)$ for $L/d_e=320$ (blue) and FFCS (red). Only the particles starting in the islands at $t/t_A=0$ are included in the $L/d_e=320$ spectra. Increasing opacity indicates later times: $t/t_A=0,1.3$ (just after $E_{R,\mathrm{max}}$), $\tau_{99}$ in blue and $t/t_A=0,5$ in red. The inset plot compares the second $L/d_e=320$ spectrum to the final FFCS spectrum (scaled to overlap), alongside a power law with index $p=3$. The green (purple) region indicates energy band  $\gamma_{\mathrm{nt}}<\gamma<\gamma_{\mathrm{hi}}$ ($\gamma >\gamma_{\mathrm{hi}}$). }
    \label{fig:spectra}
\end{figure}

\textit{Particle acceleration.}---While efficient non-thermal particle acceleration is expected to occur at the primary current sheet, the outflow-driven dynamics may activate additional acceleration mechanisms downstream. The presence of these mechanisms is apparent in the electron energy spectra shown for $L/d_e=320$ in Fig.~\ref{fig:spectra},  which develop a clear ``ankle'' feature within the non-thermal tail, near $\gamma-1=15$. For comparison, we also show the spectrum from a simulation of a force-free current sheet (FFCS) setup \footnote{This equilibrium is obtained by setting $\epsilon=0$ in Eq.~(\ref{eq:equil})  and choosing $L=6d_e$ (now the initial current sheet half-thickness) so the drift velocities  $U_e=U_i < c$. The out-of-plane magnetic field component is replaced by an external guide field $B_g/B_0=1$, but all other parameters remain the same.}  whose domain size $640d_e\times320d_e$ is comparable to the local current sheet region in the $L/d_e=320$ simulation.
Similar to previous studies, its spectrum is well described by a M-J distribution plus a power-law tail \cite{hoshino_efficiency_2022,hoshino_energy_2023,french_particle_2023,guo_particle_2015}, $f\propto (\gamma-1)^{-p}$, with a spectral index $p\sim3.3$ \cite{comisso_particle_2018,comisso_interplay_2019,french_particle_2023,vega_turbulence_2022}. Notably, as shown in the inset plot, this power law is consistent with the spectrum observed at a time just after $E_{R,\mathrm{max}}$ during island coalescence, indicating that this is the spectral signature of acceleration at the primary current sheet. As coalescence proceeds, further acceleration here is suppressed by the island bouncing, resulting in less efficient acceleration into the power-law tail. Instead, a large population of particles passing through the weakened current sheet is accelerated by additional mechanisms downstream, producing a steeper non-thermal component on top of the initial power law between $4.77<(\gamma-1)<15.40$. 

To better characterize the non-thermal component, we make use of the region decomposition to compute the energy gain by region ($\Delta\gamma_i$) for a large number of non-thermal particles ($>10^3$) which have been self-consistently tracked. Particles are separated into two energy bands which contain the steeper component below the ankle, $\gamma_{\mathrm{nt}}<\gamma<\gamma_{\mathrm{hi}}$, and the power law component above the ankle, $\gamma >\gamma_{\mathrm{hi}}$, as shown by the highlighted regions in Fig.~\ref{fig:spectra}. The non-thermal energy $\gamma_\mathrm{nt}$ is chosen as the energy where the thermal fit makes up 10\% of the total spectrum, and the high energy cutoff $\gamma_\mathrm{hi}$ is the energy halfway between the locations of the steepest and hardest spectral indices above $\gamma_\mathrm{nt}$ (see Table~\ref{tab:table1}).

\begin{figure}
    \centering
    \includegraphics[width=0.85\linewidth]{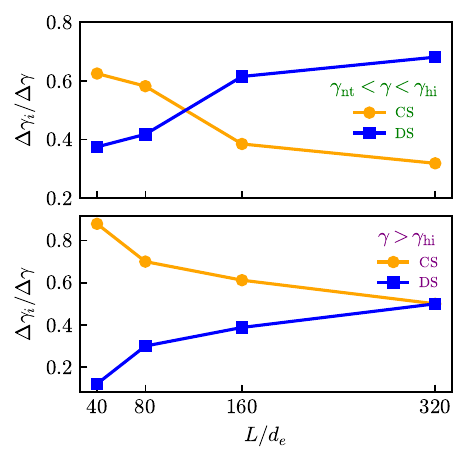}
    \caption{Fraction of energy gain by region, averaged over particles in the $\gamma_{\mathrm{nt}}<\gamma<\gamma_{\mathrm{hi}}$ (top) and $\gamma >\gamma_{\mathrm{hi}}$ (bottom) energy bands for the summed contribution of current sheet and islands (orange), and downstream region (blue).} 
    \label{fig:dgamma}
\end{figure}

The fraction of energy gained in the downstream and current sheet regions, averaged across all particles in a given band are shown in Fig~\ref{fig:dgamma}. As system size increases, there is a clear transition for $L/d_e\geq160$ in the lower energy band, showing that a dominant fraction of the energy gain can occur in the downstream region. While the highest energy particles are accelerated primarily at the current sheet for all system sizes, this band also demonstrates an increasing fraction of acceleration downstream. These results are consistent with our interpretation of the energy spectra and indicate that particle acceleration is influenced by a combination of mechanisms driven by reconnection throughout the global system; like the energy dissipation, it is de-localized from the primary reconnection site. The details of several characteristic particle trajectories for each band can be found in the Supplemental Material.

\textit{Conclusions.}---This Letter has demonstrated that the effects of multi-scale coupling play an integral role in mediating the global energy conversion and particle acceleration during magnetic reconnection. In particular, reconnection-driven turbulence and secondary current sheet formation activate sites of energy dissipation and mechanisms of non-thermal particle acceleration which are spread broadly throughout the system. As a result, they are causally connected to, but spatially and temporally decoupled from the bursty dynamics at the primary reconnecting sheet. The partition of released magnetic energy into multiple non-thermal acceleration channels gives rise to distinct features in the particle energy spectra, which differ from spectra obtained from a local current sheet setup. 

While the dominant mechanisms of particle acceleration are known to be sensitive to certain parameters, such as the guide field \cite{dahlin_mechanisms_2014,dahlin_parallel_2016} or magnetization \cite{guo_particle_2015}, and the interplay between current sheet and islands is likely sensitive to the details of the macroscopic drive \cite{granier_driven_2025}, the multi-scale features observed during island coalescence are conceivably generic to a wide range of real reconnecting systems. These results suggest that a multi-scale modeling approach is necessary for a complete understanding of the particle acceleration and energy partition in these systems.   

Authors acknowledge support from MathWorks under an MIT MathWorks Fellowship, NASA under FINESST grant No. 80NSSC24K1868 and HTMS, as well as from the DOE FES Research in Basic Plasma and Engineering program. Simulations were performed using resources at the National Energy Research Scientific Computing Center (NERSC), a Department of Energy User Facility using NERSC awards FES-ERCAP0031419 and FES-ERCAP0033013.

\bibliography{references}

\newpage

% \title{Effects of Multi-Scale Coupling on Particle Acceleration and Energy Partition in Magnetic Reconnection: Supplemental Material}% Force line breaks with \\

% \maketitle

\section*{Supplemental Material}
\section{Current Density}

Animations of the magnitude of the current density, $|J|$ for $L/d_e=40$ and $L/d_e=320$ are included with this supplementary material. At $L/d_e=40$, two plasmoids appear at the central current sheet once it reaches its maximum length, around $t/t_A=0.96$. One of these plasmoids forms closer to the central X-point and thus grows to a larger size before being ejected in the positive z-direction. As it arrives downstream, a secondary current sheet forms at it's leading edge before it is broken apart by diverging outflows. The associated fluctuations can clearly be seen propagating along the top island separatrices between $t/t_A=1.25$ and 1.5. Before and after this plasmoid is ejected, the wing regions remain laminar. As the outflows converge downstream, there is additional current sheet formation, especially as the larger fluctuations associated with the large plasmoid arrive at $t/t_A=1.5$.

At $L/d_e=320$, many more plasmoids form and are visible starting at $t/t_A=0.77$. As these plasmoids are ejected, they drive copious secondary current sheets in the wing regions, some of which are additionally plasmoid unstable. A notable example of this is at $t/t_A=1.4$, where a very large plasmoid is ejected in the negative z-direction. Due to the interactions between the outflows, plasmoids, and these secondary instabilities, the wing regions develop turbulence, contrary to the mostly laminar behavior at $L/d_e=40$. As the outflows converge further downstream, these regions also become turbulent are are filled with many secondary current sheets. This is most visible between $t/t_A=1.4$ and 2.0, associated with fluctuations driven during the first reconnection event (before the bounce). By $t/t_A=1.7$, the primary current sheet has clearly become much smaller due to the island bouncing, while the downstream turbulence remains active. Around the secondary peak in the reconnection rate, after $t/t_A\sim1.8$, several more plasmoids form at the central reconnection site and and drive similar dynamics again, albeit now at smaller scale separation due to the depletion of island flux compared to the initial condition. 

\begin{figure}
    \centering
    \includegraphics[width=0.95\linewidth]{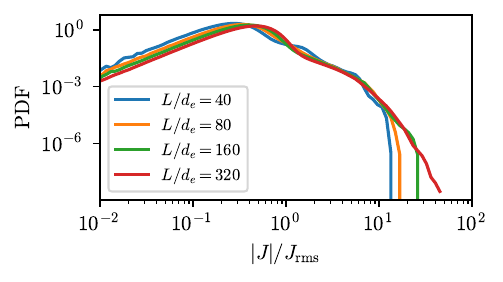}
    \caption{Probability density function of current density, normalized by the root-mean-square value at $t/t_A=1.7$ for each simulation.}
    \label{fig:jrms}
\end{figure}

Fig. \ref{fig:jrms} compares probability density functions of the magnitude of the current density normalized to its RMS value, $|J|/|J|_\mathrm{rms}$, for different $L/d_e$ at $t/t_A=1.7$. This is during an island bounce in the largest systems, where outflows have had sufficient time to propogate downstream and drive secondary current sheet formation there. As system size increases, heavier tails confirm the heightened presence of intense current sheets downstream. 

\section{Region Decomposition}

A novel feature of our setup is the separation of electron and positron species into four subspecies each according to the separating function
\[
s(x) = L\cosh^{-1}(1+\epsilon-\epsilon\cos(x/L)).
\]
At initialization, particles in the region [$-2\pi L<x<0$] ([$0<x<2\pi L$]), $|z|<s(x)$ are defined as the island 1 (island 2) species, and particles in the region $z>s(x)$ ($z<-s(x)$) are defined as the top (bottom) species. This decomposition enables detailed analysis of dynamics in different regions of the global reconnecting system. 

At each time, the islands are simply defined as the regions where the density of island particles makes up more than 99\% the total particle density, $n_{\rm{isl}}/n>0.99$. The top and bottom regions are defined similarly. As reconnection proceeds, the islands shrink as particles (and flux) from the islands pass through the current sheet and arrive downstream, resulting in a region with a mix of particle densities, which can further be divided into the current sheet and downstream regions.  

Because the length and width of the current sheet evolve dynamically with time, and may contain plasmoids much larger than the typical current sheet width $\sim d_e$, correctly identifying the current sheet region throughout the coalescence process requires a detailed algorithm. To begin, we identify the x-location of the current sheet by searching between the island centers along $z=0$ to find points where neither island region condition is met, $n_{\rm{isl}}/n<0.99$. This is not always the exact center of the box because the entire island system can be translated in the periodic x-direction due to the “pull” reconnection sites at the box edges (the initial perturbation forces the islands to coalesce by pinching at the end points, which also allows reconnection here between the flux in the top and bottom regions. We have found that these regions do not meaningfully affect the results presented here). Once the x-location of the current sheet has been identified, its vertical extent is determined by finding the point along the outflows at which a horizontal line between island boundaries reaches a maximum effective current sheet width, $\delta_{\mathrm{CS}}$. If no such point is found in either direction, the current sheet boundary is instead set by the tangent line connecting the upper boundaries of the islands, so the current sheet is limited by the total island height.

The $\delta_{\mathrm{CS}}$ parameter is set at each time step to be a fraction of the maximum height of the islands in the z-direction, $h_\mathrm{isl}$, in order to approximate the current sheet length, $l_\mathrm{CS}$. For the purposes of this paper, $\delta_{\mathrm{CS}}=0.25h_\mathrm{isl}$, so that the largest plasmoid widths, typically $\sim E_R l_{\mathrm{CS}}\sim 0.1 h_\mathrm{isl}$, are always captured. Although the separation of the system into current sheet and downstream regions always imposes hard boundaries onto a continuous domain, our choice of $\delta_{\mathrm{CS}}$ is conservative in favor of the current sheet, allowing it to extend essentially up to the island boundaries. 
For the edge case where there are no island regions found, it is also assumed that there is no current sheet. 

Finally, because we have found that the regions of high energy dissipation are slightly wider than the current sheet itself (non-ideal effects captured by the electron-frame dissipation measure can be
strong just upstream in the islands), we increase the current sheet width by 2$d_e$ in both the positive and negative x-directions to include these regions in the current sheet contribution. 

After identifying the top, bottom, island, and current sheet regions, the reconnected region, which contains all of the reconnected flux, is simply taken to be all of the remaining area in the simulation domain. A movie showing this decomposition at each time for $L/d_e=320$ is included with this supplementary material. 

\section{Particle Trajectories}

We present characteristic particle trajectories for each energy band. We note that these represent only a small sample of a diverse set of trajectories. 

\subsection{Intermediate (non-thermal) energy band: $\gamma_\mathrm{nt}<\gamma<\gamma_\mathrm{hi}$}

\begin{figure*}
    \includegraphics[width=0.75\linewidth]{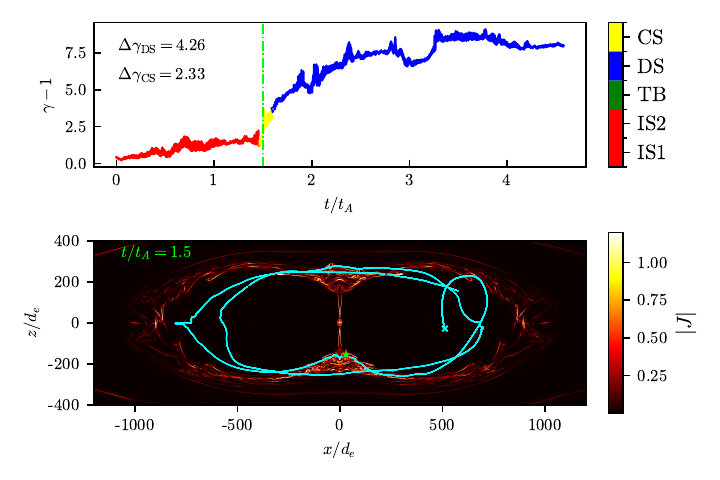}
    \caption{Particle in the energy interval $\gamma_\mathrm{nt}<\gamma<\gamma_\mathrm{hi}$ which gains energy in both the current sheet and downstream regions. 
    The upper panel shows kinetic energy versus time, color coded by region assigned to particle at each time. Current sheet is shown in yellow, downstream in blue, top and bottom in green, and islands in red. The energy gained by the particle in the current sheet and downstream regions, $\Delta\gamma_\mathrm{CS}$ and $\Delta\gamma_\mathrm{DS}$ are shown in the upper left corner. 
    The lower plot shows the particle trajectory (cyan) overlaid with current density magnitude $|J|$ at $t/t_A=1.5$ . The particle position at $t/t_A=0$ is given by the cyan `X' and the position at the time of the current density plot is given by the magenta triangle (also marked by vertical line in upper plot). }
    \label{fig:nt_traj_csds}
\end{figure*}

This energy band represents particles whose total energy at $\tau_{99}$ are within the non-thermal distribution, ($\gamma>\gamma_{\rm{nt}}$), but not a part of the harder power law component, $\gamma<\gamma_{\rm{hi}}$. Figure~\ref{fig:nt_traj_csds} shows the trajectory of a particle in this band which gains energy at the current sheet and downstream. The upper plot shows the kinetic energy, $\gamma-1$, against time and is color coded by region. The particle arrives at the current sheet at $t/t_A\approx1.5$ and receives a small increase to its energy. As it travels through the downstream region, it gradually gains additional energy. The trajectory is overplotted with current density in the lower plot, with the start time marked by a cyan `X' and the time of the current density background plotted marked with a green star. This time is also marked by a vertical green line in the upper plot. 

\begin{figure*}
    \includegraphics[width=0.75\linewidth]{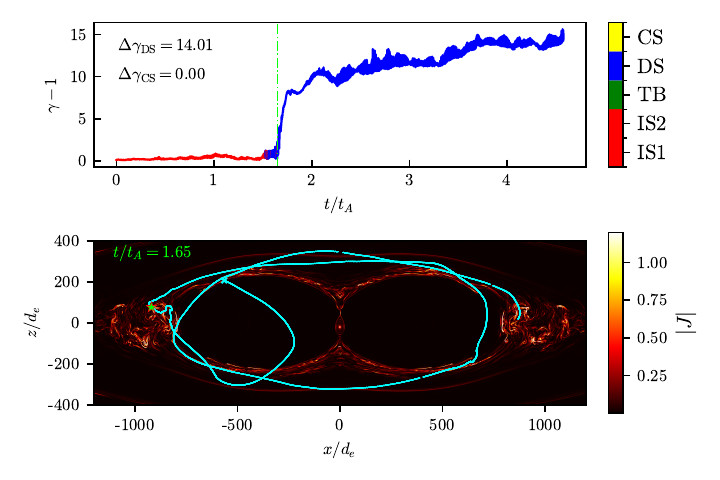}
    \caption{Similar to Fig.~\ref{fig:nt_traj_csds} for a particle in the energy interval $\gamma_\mathrm{nt}<\gamma<\gamma_\mathrm{hi}$ which gains energy primarily downstream.}
    \label{fig:nt_traj_ds}
\end{figure*}

Figure~\ref{fig:nt_traj_ds} shows the trajectory of a particle in the same final energy interval but which gains most of its energy downstream. Although it starts in the island, it reaches the downstream region without passing through the current sheet, as it eventually reaches the island edge and is picked up by the downstream flows. It then experiences a rapid increase in kinetic energy as it interacts with a secondary current sheet in this region. 

\subsection{High energy band: $\gamma>\gamma_\mathrm{hi}$}

\begin{figure*}
    \includegraphics[width=0.75\linewidth]{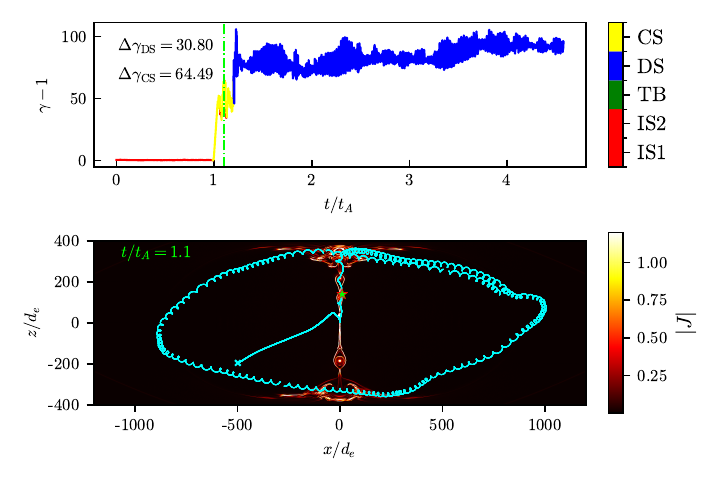}
    \caption{Similar to Fig.~\ref{fig:nt_traj_csds} for a particle in the energy interval $\gamma>\gamma_\mathrm{hi}$ which gains energy primarily at the current sheet.}
    \label{fig:tail_traj_cs}
\end{figure*}

This energy band represents particles whose total energy at $\tau_{99}$ is within the harder power law component, $\gamma>\gamma_\mathrm{hi}$. Figure~\ref{fig:tail_traj_cs} shows the trajectory of a particle in this band which gains most of it energy at the current sheet. As expected, it directly samples the reconnection x-point, where it rapidly gains a significant amount of energy. The particle is trapped in a plasmoid and is eventually ejected into the downstream, where it first experiences another period of rapid energy gain, before continuing to be accelerated over a longer timescale. 

\begin{figure*}
    \includegraphics[width=0.75\linewidth]{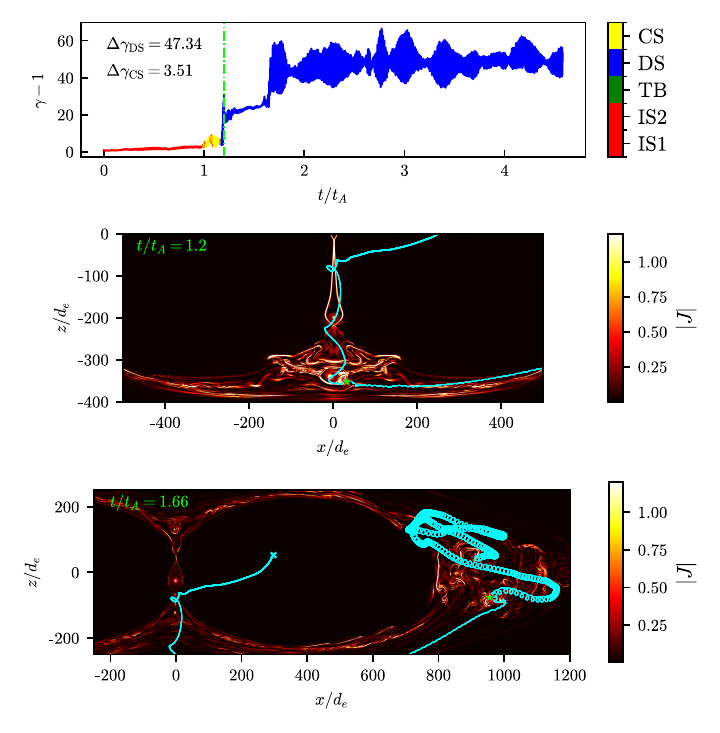}
    \caption{Similar to Fig.~\ref{fig:nt_traj_csds} for a particle in the energy interval $\gamma>\gamma_\mathrm{hi}$ which gains energy primarily in the downstream region. The center and lower plots show the particle trajectory and current density at two different times where there is intense particle acceleration.}
    \label{fig:tail_traj_ds}
\end{figure*}

Figure~\ref{fig:tail_traj_ds} shows the trajectory of a particle which gains most of its energy downstream despite falling within the high energy band. The upper plot shows that there are two distinct acceleration events at $t/t_A=1.2$ and 1.6 where there is a rapid increase in the kinetic energy. Notably, both of these occur after the particle has transited the current sheet, where it does not undergo significant acceleration. The center plot shows that the first of these is associated with a secondary current sheet (which has produced an additional plasmoid) forming in the wing region. The bottom plot shows that the later acceleration can be attributed to the interaction with a current sheet in the downstream turbulence.  

\end{document}